\def\cmv {cm$^{-3}$}

\def\to {$\rightarrow$}
\def\as {\arcsec}
\def\water {H$_2$O}
\def\chisq {$\chi^2$}
\def\tk {T$_K$}
\def\kms {km s$^{-1}$}
\def\meann {$\langle$log(n)$\rangle$}
\def\meancd {$\langle$log(N)$\rangle$}
\def\lcs {$L ({\rm CS \ 5-4})$}
\documentstyle[aaspp]{article}
\begin{document}
\title{Dense gas and Star Formation: \\
Characteristics of Cloud Cores Associated with Water Masers. }
\author{Ren\'e Plume, D.T. Jaffe, Neal J. Evans II}
\affil{Department of Astronomy, University of Texas at Austin}
\authoraddr{Austin, Texas 78712-1083, U.S.A.}
\author{J. Mart\'in-Pintado, and J. G\'omez-Gonz\'alez}
\affil{Centro Astronomico de Yebes}
\authoraddr{Guadalajara, Spain}
\begin{abstract}

We have observed 150 regions of massive star formation, selected
originally by the presence of an \water\ maser,
in the J = 5\to4, 3\to2, and 2\to1 transitions of CS,
and 49 regions in the same transitions of C$^{34}$S.
Over 90\% of the 150 regions were detected in the J = 2\to1 and 3\to2
transitions of CS and 75\% were detected in the J=5\to4 transition.
We have combined the data with
the J = 7\to6 data from our original survey (Plume et al. 1992)
to determine the density by analyzing the excitation of the rotational
levels.  Using Large Velocity Gradient (LVG) models,
we have determined densities and column densities for 71 of these
regions. The gas densities are very high (\meann\ = 5.9), but
much less than the critical density of the J=7\to6 line.
Small maps of 25 of the sources in the J = 5\to4 line yield
a mean diameter of 1.0 pc. Several estimates of the mass of dense gas
were made for the sources for which we had sufficient information. The mean
virial mass is 3800 M$_{\sun}$. The mean ratio of bolometric
luminosity to virial mass ($L/M$) is 190, about 50 times higher than
estimates using CO emission, suggesting that star formation is much more
efficient in the dense gas probed in this study.
The gas depletion time for the dense gas is roughly $1.3 \times 10^7$ yr,
comparable to the timescale for gas dispersal around open clusters
and OB associations.  We find no statistically significant
linewidth--size or density--size relationships in our data.
Instead, both linewidth and density are larger for a given size than
would be predicted by the usual relationships.
We find that the linewidth {\bf increases} with density, the
opposite of what would be predicted by the usual arguments.
We estimate that the luminosity of our Galaxy (excluding the inner
400 pc) in the
CS J = 5\to4 transition is 15 to 23 L$_{\sun}$, considerably less
than the luminosity in this line within the central 100 pc of
NGC 253 and M82. In addition, the ratio of far-infrared luminosity to
CS luminosity is higher in M82 than in any cloud in our sample.

\end{abstract}
\section {Introduction}

Very dense gas (n $\geq 10^5$ \cmv) has an important
effect upon star formation in molecular clouds.
The presence of very dense gas
affects the Jeans mass and other measures of stability.
In addition,
the quantity of very dense gas has consequences for
the calculated star formation efficiency, since it is this
material that
actively participates in star formation  (\markcite{Lada et al. 1991};
\markcite{Solomon, Radford, \& Downes 1990}).
Most stars (even low-mass stars) form in regions where high-mass
stars are forming (Elmegreen 1985).
In high-mass star-forming regions,
winds and radiation from nearby, newly-formed stars can disrupt
the local gas  and effectively shut down further star formation.
Cores of very dense gas, however,
resist these disruptive forces and can
help to maintain
star formation in the hostile environments associated with young, massive
stars
(\markcite{Klein et al. 1983};
\markcite{LaRosa 1983}).

How universal is very dense gas (n $\geq 10^5$ \cmv) ?
\markcite{Benson \& Myers (1989)} and \markcite{Zhou et al. (1989)}
have shown that densities of order 10$^4$ to $10^5$ \cmv\ are common in
low-mass star-forming regions.
Studies of a few selected regions that are forming massive stars
(e.g., \markcite{Jaffe et al. 1983};
\markcite{Cunningham et al. 1984};
\markcite{Snell et al. 1984};
\markcite{Richardson et al. 1985};
\markcite{Mundy et al. 1987};
\markcite{Mezger et al. 1988};
\markcite{Churchwell, Walmsley, \& Wood 1992};
\markcite{Wang et al. 1993};
\markcite{Bergin et al. 1996};
Hofner et al. 1996)
 have demonstrated the presence very dense gas (n $= 10^5$ to $10^6$ \cmv ).
However, it is unclear whether such gas is common to all regions forming
massive stars. The overall sample of such regions is
small, and the studies
used a variety of  selection criteria and density measurement techniques.

To assess the prevalence of very dense gas,
we need to determine densities using a consistent method
in a large and representative sample
of regions forming massive stars.  The
small beam sizes at frequencies used to probe for very dense gas,
along with limited amounts of available telescope time, make it
impossible to map completely all the regions known to be forming
massive stars.  Therefore, we require
a pointer to likely locations of active star formation within
molecular clouds.   H$_2$O masers are ideal
for this purpose since they contain at least small amounts of
extremely dense gas (n $> 10^{10}$ \cmv)
(\markcite{Elitzur et al. 1989}; \markcite{Strelnitskij 1984}); and, in
well-studied regions, they are intimately associated with
star formation (e.g., \markcite{Genzel \& Downes 1977, 1979};
\markcite{Jaffe et al. 1981};
\markcite{Wood \& Churchwell 1989};
\markcite{Churchwell 1990}).

Our initial survey (\markcite{Plume et al. 1992}; hereafter Paper I)
searched for thermal emission from dense gas associated with
H$_2$O masers.
The sample consisted of 179 of the 181 H$_2$O masers listed in the
catalog of \markcite{Cesaroni et al. (1988)} as ``HII region''
(i.e., not a late-type star) masers which were north of $\delta=-30$\deg\
and had positions known to better than 8\arcsec.
We used the J = 7$\rightarrow$6 transition of CS, which has a high
critical density (n$_{crit} \approx 2\times10^7$ \cmv),
selecting regions of very dense gas.

Observations of a single transition of the CS molecule are not sufficient to
determine gas density.
Therefore, we have observed
the J = 5$\rightarrow$4, 3$\rightarrow$2, and 2$\rightarrow$1 transitions
of CS in 150 of the 179 regions from
our initial CS J = 7$\rightarrow$6 survey (Paper I). Of the
regions sampled in the current survey, 85 (57\%) contained
CS 7$\rightarrow$6 emission; this percentage is similar to that of
the original CS 7\to6 survey (104 of 179, or 58\%; Paper I),
indicating that this
study was not biased towards the densest regions.
We also observed the same transitions of C$^{34}$S in 49 of
the strongest CS sources.
The C$^{34}$S data yield an independent measurement of the densities.
We have also mapped 21 sources in the J = 5\to4 line of
CS to determine the size of the cores.

In \S 3,  we present basic detection statistics
and discuss the individual spectra and the maps.
In \S 4, we present densities and column densities
based upon excitation analysis of the data, consider effects of
opacity and temperature uncertainties on the results, and compute
masses. In \S 5, we consider issues like the star formation efficiency,
compare these regions to other regions, and estimate the luminosity
of the galaxy in the CS J = 5$\rightarrow$4 line.

\section {Observations}

We observed the J = 5\to4, 3\to2 and 2\to1 transitions of CS and C$^{34}$S in
1990 June, 1991 April, and 1991 October at the IRAM 30-m telescope at Pico
Veleta, Spain, with the 3-mm, 2-mm, and 1.3-mm SIS
receivers  tuned to single-sideband mode.
Table 1 lists the line frequencies,
main beam efficiencies ($\eta_{mb}$), beamsizes, typical system
temperatures, and velocity resolutions for each transition.
To convert to the T$_R^*$
scale \markcite{(Kutner and Ulich 1981)}, the data were scaled by $\eta_{mb}$
(i.e., T$_R^*$ = T$_A^*$/$\eta_{mb}$).
In IRAM notation, $\eta_{mb} = {B_{eff}}/{F_{eff}}$, the
back spillover and scattering efficiency divided by the forward spillover
and scattering efficiency.
For all excitation calculations we have assumed that the source is
fully resolved, so that T$_R^*$ = T$_R$,
the Rayleigh-Jeans
temperature of a spatially resolved source observed with a perfect
telescope above the atmosphere.


The pointing was checked regularly. In 1990 June, we pointed on
continuum emission from K3-50, W3(OH), and Jupiter and found
the pointing to be
very sensitive to changes in azimuth and elevation, with a
maximum spread of $\approx$ 15\as.  Consequently, we made nine-point
maps on a 16\as\ grid for each source, while defining a pointing
curve. The reliability of these observations will be discussed below.
The pointing curves were well-determined by the time we made the C$^{34}$S
observations.
In 1991 April, we checked the pointing with continuum observations of
W3(OH), NGC 7027, BL Lac, and Saturn.
For this observing run,
the telescope pointing was accurate to within 4\as\ rms.
In 1991 October, the pointing was good to 5\as\ rms.

The data were calibrated to the T$^*_A$ scale using the chopper
wheel method (\markcite{Penzias \& Burrus 1973}).  To include the effects of
different sideband gains, we observed the line calibration sources
W51M, W51N, W44, W3(OH), and DR21S in 1990 June, and
IRC+10216, W3(OH), and Orion IRc2 in 1991 April.
These sources were observed by \markcite{Mauersberger et al. (1989)}
in single-sideband mode with an image-sideband rejection of $>$ 8 dB.
We compared our observed antenna temperatures
with those tabulated  by \markcite{Mauersberger et al. (1989)} and adjusted our
temperature scale to agree with theirs.
For the 1990 June run, we did not need to adjust the
CS J = 2\to1 data.  The June
CS 3\to2 and 5\to4 transitions needed to be
reduced in strength by $\approx$ 10\%.
The 1991 April CS J = 2\to1 and 5\to4 data both needed to be increased
by 20\%.   The 1991 April CS J = 3\to2
data were multiplied by a factor of 2 and assigned a 50\% calibration error.
Lacking a standard source for the C$^{34}$S J = 3\to2 line, we also scaled it
up by a factor of 2 and assigned a 50\% calibration error.
In 1991 October, no scaling was necessary.
We have also compared the CS J=7\to6 results of
\markcite{Plume et al. (1991)} to the more recent single-sideband observations
of \markcite{Wang et al. (1994)}.  For the five sources in common,
the ratio of T$_A^*$ values is 0.98$\pm$0.19, indicating that the calibration
of the J = 7\to6 data in Paper I was good.

We also used the CSO for some auxiliary observations, with parameters shown
in the last 4 lines of Table 1.
We made small maps of the CS J = 5\to4 transition
toward 21 of the sources in our sample in 1994 June.
The pointing was checked on planets and repeated to 3\as.
We observed the J = 7\to6 transition of C$^{34}$S in 1991 June.
Beam sizes and efficiencies are  based on Mangum (1993).
For a few sources, we also observed the J = 10\to9 transition of CS in 1993
December. The pointing was checked on planets and found
to be constant to 6\as.
The beamsize was assumed to be 14\as, based on scaling from
lower frequencies, and the efficiency was measured on Saturn.
Finally, the J = 14\to13 line of CS was observed toward two sources in
1994 December. The pointing was checked by observing planets
and was constant to 6\as. The beamsize was assumed to be 11\as.
The optical depth at the zenith was measured by tipping to be 0.35 at
685 GHz.
For all the CSO observations, two AOS spectrometers were used, with
resolutions of about 140 kHz and 1 MHz. The choice of spectrometer for
determining line parameters was made on the basis of getting sufficient
resolution and signal-to-noise in the line.

\section {Results}

\subsection{Detection Statistics}

Table 2 lists the 150 H$_2$O maser sites
observed at IRAM in the CS J = 5\to4, 3\to2, and 2\to1 transitions.
Tables 1 and 2 of Paper I list the positions of the masers and
the CS J=7\to6 line parameters or upper limits.
Table 2 of the present paper lists the source names in order of
increasing galactic longitude,
the radiation temperature (T$_R^*$), integrated
intensity ($\int{T_R^*dv}$), velocity centroid (V$_{LSR}$), and
full width at half maximum (FWHM) for the three transitions.
For CS data obtained in 1990 June, we list the line parameters at the
position in the nine-point map with the strongest emission in the
J=5\to4 line. This choice is based on the results of \S 3.2, where we
find that the J=5\to4 emission almost always peaks at the maser position.
While the line parameters for 1990 June are useful in detection statistics and
as a guide for follow-up work, we have found that the position correction
was inadequate for them to be used together with the J=7\to6 data to
determine densities; therefore we do not use the 1990 June data in \S 4.
For undetected lines, the upper limits to T$_R^*$ are 3$\sigma$.
For CS J = 3\to2 and 2\to1,
we have tabulated only the data with the highest
spectral resolution.   We also observed the C$^{34}$S lines in 49 of
the strongest CS emitters.
The results for C$^{34}$S are presented in Table 3. Transitions
listed with dashes (--) instead of values or
upper limits to T$_R^*$ were not observed.
Table 4 has the results for J=10\to9 and 14\to13.

Usually, we obtained the line parameters from Gaussian
fits to the lines but
some sources listed in Table 2  had spectra with more
than one peak.
To determine the line parameters in these cases,
we took the following approach.
First, if  the profiles of the higher J (i.e., 7\to6 or 5\to4) lines or
C$^{34}$S
lines (where available)
matched one or more of the peaks seen in the lower
J transitions, we assumed that
the source was composed of distinct cloud
components (e.g., Figure 1a); and we derived
the line parameters by performing a multiple
Gaussian fit to the whole profile.
Each Gaussian component is listed individually in Table 2.
Three sources have 2 velocity components and one has 3 components; these
are identified in Tables 2 and 3 by  the notation `` C\#''  (where
\# is the component number).  With the inclusion
of all the separate components, Table 2 displays results for 155
cloud components.
Second, if comparison of CS data with
C$^{34}$S data indicated that the
CS line was self-absorbed (Figure 1b shows an example of this situation),
we calculated the line parameters ($\int T_R^* dV$, V$_{LSR}$, and FWHM)
from moment integrals over the profile. Then $T_R^*$ was calculated from
$\int T_R^* dV$ /FWHM (values given in parentheses in Table 2).
Only 18 of the  150 spectra
were obviously self-absorbed in CS 2\to1, with smaller numbers showing obvious
self-absorption in the higher-J lines.
Of course, self-absorption may exist at a less obvious level in other sources.

Figure 2 illustrates the detection rate for the observed CS transitions.
The distribution counts as detected
only those sources with observed T$_R^* \geq 0.5$K .
Because the sensitivity achieved for
the CS J = 7\to6 line (Paper I) was similar to that for the lower
J transitions, the drop in the detection rate towards higher rotational
levels reflects a real drop in the number of
sources exhibiting emission at the same level in the higher J lines.

\subsection{Extent of the Dense Gas: CS J = 5\to4 Maps}

To determine the effect that very dense gas has upon star formation,
we need to know the extent of the gas and its location within the star-forming
regions. We have observed 21 of our sources in the
CS 5\to4 line with the CSO. For each source, we made a cross-scan in R.A.
and Dec., typically consisting of 9 points. For most of the sources, the
separation of the observed points was 30\as.  For a few of the smaller
sources, we made the observations at 15\as\ intervals.
In addition,  we have assembled from the literature
data taken with the same equipment for four
other sources from our survey.
Table 5 lists the mapping results
for all 25 sources.  The integrated intensities
listed in Table 5 are for the interpolated maximum along each cross scan.
{}From the maps we derived diameters and beam correction factors,
$F_c = (\Omega_{source}+\Omega_{beam}$)/$\Omega_{beam}$.
The beam correction factors were calculated assuming that
a Gaussian was a good representation of both the beam shape and the source
intensity distribution.  Using the integrated intensity, the $F_c$, and
the distances, $d$(kpc), we calculated the luminosity in the CS J=5\to4
line from
\begin{equation}
 L({\rm CS \ 5-4}) = 1.05 \times 10^{-5} L_{\sun} d^2 F_c  \int{T_R^*dv}.
\end{equation}

Table 5 also lists the offsets of the CS 5\to4 peaks from the maser
positions in arcseconds.
With the exception of a few of the larger sources, differences
in the peak position of
the CS 5\to4 distribution and the H$_2$O maser position are smaller than
the combined pointing uncertainties and maser positional uncertainties
($\pm$3\as\ and $\leq\pm$8\as, respectively).
Jenness et al. (1995) have also found a very good correlation between the
peak of the submillimeter emission and the maser position.
The mean diameter of the sources listed in Table 3 is 1.0 pc.
The dispersion about this mean, however, is large (0.7 pc).  If one
examines sources at $d\leq 3.0$ pc, the mean diameter is 0.5 pc with
a dispersion of 0.4 pc.  This difference, while significant, probably
does not arise from observational biases in the CS data.  Most
of the more distant sources are well resolved and bright.  It is more
likely that the differences arise from selection biases in the original
samples used to search for H$_2$O masers.
Complete mapping of the CS 5\to4 line in several sources gives similar
sizes. The emission in NGC2024 has a diameter of 0.4 pc, while S140 has
a diameter of 0.8 pc (Snell et al. 1984). The emission in M17 is more
extensive:
2.3 pc in 5\to4 (Snell et al.); 2.1 pc in 7\to6 (Wang et al. 1993).

\section{Analysis}

With the addition of the lower J transitions in the present study to the
CS J = 7\to6 data from Paper I, we can
determine densities in a large sample of star-forming regions.
In \S 4.1, we discuss the calculations and examine the effects
of  opacity and uncertainties in kinetic temperature
on density and column density determinations.
In \S 4.2, we consider the effects of density inhomogeneities, and
we compute masses in \S 4.3.

\subsection{Densities and Column Densities}

To determine densities and column densities,
we used a large velocity gradient (LVG) code to solve
the coupled equations of statistical equilibrium and radiative transfer,
including the first 20 rotational levels of CS in the calculation.
We assume that the gas has a constant density and temperature and that it
uniformly fills all the beams used in this study.
We calculated a 20$\times$20 grid of radiation temperatures
in column density per velocity interval -- density space
for a kinetic temperature of 50 K.
The CS densities in the LVG model grid ran from
$10^4$ to $10^8$ \cmv, and  the column densities per velocity interval
(N/$\Delta$v) ranged from $10^{11}$ to $10^{16}$
cm$^{-2}$/km s$^{-1}$.   These ranges span the parameter space of
all solutions which fit our data.  All the models converged to a solution.

Using a $\chi^2$ minimization routine, we fit the LVG models to the
observed CS line intensities.
Table 6 lists the densities for 71 sources.
We have not included fits for the CS data obtained in 1990 June, for reasons
discussed below. We have listed the log of the density and column density,
along with the value of $\chi^2$ and a key to which transitions were used
and whether the lines were self-absorbed. The values of density and column
density apply to the region selected by the typical beam used for
the observations (about 20\arcsec ). The $\chi^2$ values allow us
to assess whether the models (any of the points in the LVG grid) are
a good representation of the data.
The distribution of $\chi^2$ values for sources with 4 transitions
(40 sources) is similar to
what is expected theoretically if the model is a reasonable fit to the data,
as is the distribution for sources with only three transitions (31 sources).
These facts suggest that our estimates of the calibration uncertainties
are reasonable. We originally included the 1990 June data in the fits, but
they had a very high percentage of bad fits, leading us to conclude that
the uncertain pointing made them unsuitable for combining with the
CSO J=7\to6 data. The 8 self-absorbed sources with fits in Table 6
(marked by a flag) do not have \chisq\ significantly worse than the other
sources. One source with 3 transitions (212.25-1.10) produced a very uncertain
density, and we have excluded it from the statistics that follow.

The mean logarithmic density for sources with detected emission from
all 4 CS transitions is \meann $= 5.93 \pm 0.23$, where 0.23 represents
the standard deviation of the distribution.
The mean logarithmic column density is \meancd $= 14.42 \pm 0.49$.
The results for the sources undetected in J=7\to6 are
\meann $= 5.59 \pm 0.39$; \meancd $ = 13.57 \pm 0.35$.
Figure 3 shows histograms of the densities and column densities.
The solid line plots the densities determined from
all 4 CS transitions and the dashed line is the density distribution
for sources without J= 7\to6 detections.
These results show that the difference between a CS 7\to6 detection
and non-detection is more related to column density than to volume density.
Therefore, the detectability of lines of
high critical density is more affected by the quantity of dense gas
present than by its density.
To check whether the difference was solely a result of having a J=7\to6 line
to fit, we re-fit the sources with 7\to6 detections,
forcing the $\chi ^2$ fitting routine
to ignore the CS 7\to6 line and to fit only the 3 lower transitions.
The resulting \meann\ is $5.71 \pm 0.19$, and \meancd\ is $14.36 \pm 0.49$.
This result confirms our conclusion that the most significant difference
between
a J=7\to6 detection and a non-detection is the column density.

What effect would high opacity in the CS lines have on the
derived densities and column densities?
Eighteen of the sources in this survey have noticeable self-absorption in
at least one transition.
In addition, an LVG model run for the mean density, column density, and
linewidth results in CS line opacities that are roughly unity.
Thus, self-absorption may affect the fits, even if it is not apparent
in the line profiles.
Since the C$^{34}$S  transitions will usually be optically thin,
we independently fit the C$^{34}$S transitions to an LVG model grid,
with a range of parameters identical to those used in the original CS grid.
Table 6 lists the densities, column densities, and $\chi{^2}$  derived from
fits to the C$^{34}$S data.
Problems with the receivers during the C$^{34}$S observations meant that
we have various combinations of lines to fit, as indicated by
the key in Table 6.
There are few sources with both adequate CS and acceptable C$^{34}$S data.
The fits to the sources with three transitions of C$^{34}$S give
\meann $= 5.95 \pm 0.20$,
essentially identical to the \meann\ derived from 4 transitions of CS.
The mean difference between CS and C$^{34}$S in log(n) is $0.07\pm0.24$,
indicating no significant difference in the derived densities.
It is unlikely that the densities calculated for sources in our survey from the
CS lines alone are seriously affected by CS optical depth.
The average isotope ratio, $N(CS)/ N(C^{34}S)$, is $5.1\pm 2.2$,
clearly less than the terrestrial ratio, and
lower than the isotope ratios of 9--17 found by Mundy et al. (1986)
and 13 (Wang et al. 1993). Chin et al. (1996) have recently found
evidence for low values of this ratio in the inner Galaxy, but our
values are lower still. It is likely that our procedure has underestimated
$N(CS)$ to some extent. For this reason, and also because
these ratios are not very well determined
for individual sources, we have adopted an isotopic abundance
ratio of 10 in what follows.

By increasing the number of transitions,
simultaneous fitting of the CS and C$^{34}$S data should,
in principle, allow us to determine the
densities and column densities more accurately.
Using the LVG model grid for CS and constraining the isotope ratio to be
10, we fit CS and C$^{34}$S  transitions simultaneously.
The results are listed in  Table 6.
While neither the densities nor the column densities
are significantly different from those determined from
fits to the CS data alone, $\chi ^2$ is considerably larger.
The poor fits probably result from assuming a fixed isotopic
abundance ratio for all sources.

It is likely that many of the regions of massive star formation
contained within this study have temperatures in excess of 50 K.
At the densities
implied by the CS observations, the gas kinetic temperature will be coupled
to the dust temperature. For grains with opacity decreasing linearly
with wavelength, one can write
\begin{equation}
T_D = C [{{L}\over{\theta^2 d^2}}]^{0.2},
\end{equation}
where $L$ is the luminosity in solar units, $d$ is the distance in kpc,
and $\theta$ is the angular separation from the heating source in arcseconds.
Using these units, $C = 15$ to 40 (Makinen et al. 1985, Butner et al. 1990).
We can estimate the range of temperatures in our sources from the
luminosities in Table 7 and distances in Table 5; $\langle (L/d^2)^{0.2}
\rangle
= 7.5\pm 1.6$. At a radius of 10$\arcsec$, characteristic of the beams in
Table 1 and the beam of the J = 7\to6 observations, $T_D = 50$ to 100 K.
To assess the effects of temperature uncertainties on the derived
source properties, we also fit the sources with 4 transitions to a grid
of models run for a temperature of 100 K. The value of \meann\ decreased by
0.3 and the value of \meancd\ was essentially unchanged. Regardless of the
assumed temperature, our data imply a thermal pressure, $nT \sim 4 \times 10^7$
K cm$^{-3}$, which is much higher than found in regions not forming
massive stars.

Within the limitations of a single-density model, we conclude that
the effects of opacity and temperature on the determinations of density
are not severe (about at the factor of 2 level). Typical densities in regions
detected in the J=7\to6 survey are $10^6$ \cmv. Toward water masers not
detected in the J=7\to6 survey, the densities are about a factor of 2 less,
but the column densities of CS are about a factor of 7 less, on average,
than the values found for regions detected in the J=7\to6 line.
The densities for both groups of sources are considerably
less than the critical density of the
CS J=7\to6 line ($2 \times 10^7$ \cmv), reminding us that detection
of emission from a hard-to-excite line does not imply the existence of gas at
the critical density. Molecules can emit significantly in high-J transitions
with critical densities considerably above the actual density because
of trapping and multilevel effects (see also Evans 1989).
For example, levels with J$>>0$ have many possible routes for excitation
by collisions, but only one radiative decay path.

The high densities found in this survey of regions forming massive stars
are similar to those obtained from other, more detailed, studies
of individual, luminous, star-forming regions (see ref. in \S 1).
Consequently, the results found from studies of a few clouds can be applied,
in a statistical sense, to the broader sample of massive star-forming
regions.

\subsection{ Multiple Density Models}

Our LVG analysis assumes that the density is uniform and that
the emitting gas fills the beam.
How good are these assumptions? Figure 4 gives examples of LVG
model fits to several of the sources: three with good fits and three
with bad fits, as measured by the \chisq\ value.
While the LVG models generally fit the data within the uncertainties,
a closer look reveals that the discrepancies between model and observation
are very consistent, even for the good fits.
Almost all fits overpredict the 3\to2 and 5\to4 lines
and underpredict the 2\to1 and 7\to6 lines.
Thus, the data have, on average, a smaller variation of intensity with
J than do the best-fit
LVG models, as would be expected for a source with
a mixture of gas at different densities.
In this section, we examine models with varying densities to
see how well they explain the intensity versus J distribution.

\markcite{Snell et al. (1984) } and Wang et al. (1993) have discussed
the effects of fitting a single density to the CS emission from
a mixture of gas at about $10^6$ \cmv\ and gas lower in density by about a
factor of 10.  They showed that, until the filling factor of the high
density gas becomes very small (i.e., $f< 0.2$), the density derived
from fitting a single density model matches that of the high density
component to within a factor of two.
The CS transitions we have observed should behave
in a similar way in that they are biased toward measuring gas with
densities close to $10^6$ \cmv.

We now ask a more radical question. Could the apparent density near
$10^6$ \cmv\ be an artifact of fitting to a single density a mixture of
ultra-dense gas (n = $10^8$ \cmv) and gas at a much lower (n = $10^4$ \cmv)
density?
In this picture, the histogram of densities (Figure 3) would be produced by
varying the filling factor of the dense component.
We chose a value of $10^8$ \cmv\ for the
density of the ultra-dense gas because the 7\to6 transition
becomes completely thermalized at that density. Thus, the component
with n$= 10^8$ \cmv\ represents any gas with n$\geq 10^8$ \cmv.
We synthesized clouds from a mixture of these two components
at 20 values of N/$\Delta$v between
$10^{12}$ and 10$^{16}$ cm$^{-2}$/km s$^{-1}$.
For each density and column density, we used the LVG code to calculate the
expected emission. We then
varied the filling factor of the ultra-dense gas ($f$) and
the low-density gas ($1-f$), with $0 \leq f \leq 1$ in steps of 0.05,
and summed the contributions to each transition for each possible combination
of $f$, column density of the gas at n$= 10^4$ \cmv\ (N$_{\rm low}$), and
column density of the gas at n$= 10^8$ \cmv\ (N$_{\rm high}$).
These results then formed
a grid of models which could be fitted to the data, just as the single-density
models had been fitted. We found that the \chisq\ value worsened,
despite the extra free parameter, for sources where the single-density
fit had been good (\chisq $\leq 1$). On the other hand, the sources
which were poorly fitted (\chisq $> 1$) with the single-density model
were better fitted with the two-density model. The two-density fits
typically required very high column densities (\meancd $= 16.16$) of
the low-density gas compared to those of the ultra-dense gas
(\meancd $= 13.85$).

To see if we could constrain the amount of ultra-dense gas in the sources
with poor single-density fits, we followed a similar, but
less restrictive, procedure.
We started by assuming that the CS J = 2\to1 and 3\to2
transitions effectively probe the low density gas in the
beam, and we used them to fit the density (n$_{\rm low}$)
and column density (N$_{\rm low}$) of the
low-density component. We then used the LVG code
to obtain  the expected emission from each rotational
transition for a gas at this density and column density at a
temperature of 50K.  These intensities, multiplied by ($1-f$),
were used to represent the
lower density component.  We then searched a parameter space of $f$ and
log(N/$\Delta$v) for the best values for the ultra-dense component
(density once again fixed at 10$^8$ cm$^{-3}$).  We summed ($1-f$) times
the lower density intensities and $f$ times the ultra-dense gas intensities
and compared this sum to the observations.
This method has a large number of free parameters: $f$, n$_{\rm low}$,
N$_{\rm low}/ \Delta$v, and N$_{\rm high}/ \Delta$v, which are constrained
by only 4 transitions. Furthermore, it
does not correct the properties of the lower
density component for the contributions of the high
density gas to the J = 2\to1 and 3\to2 emission. We use it for
illustrative purposes only. We show the two-density fits as dashed
lines in Figure 4, but we do not tabulate the results. The mean
properties of these solutions for the sources with single-density \chisq $> 1$
are as follows: $f = 0.22$, log(n$_{\rm low})
= 5.4 \pm 0.3$, log(N$_{\rm low}) =
14.39$, and log(N$_{\rm high}) = 14.39$ (equal column densities in the two
components). Thus,
in general, the filling factor of ultra-dense gas is small (less than
25\%), and the data still favor a large amount of gas at $n > 10^5$ \cmv.

Another possible source model is a continous density gradient, such
as a power law. Power-law density distributions have been proposed for
regions of low-mass star formation on theoretical grounds (Shu 1977)
and seem to fit the observations well in some cases (e.g., Zhou et al. 1991).
They have also been applied to some regions forming stars of higher mass
(e.g., Zhou et al. 1994; Carr et al. 1995). The latter reference is
particularly relevant here, as it included a more complete
analysis of GL2591 (called CRL2591 in this paper),
including data from this paper, but adding other data. While Table 6
indicates a good fit to the data for that source with a single-density model,
Carr et al. found that a single density cannot fit all the data,
when other data are included, particularly J = 5\to4
and 10\to9 data from the CSO.
They developed models with power-law density and temperature gradients
that fit all the data. We can use the example of CRL2591 to explore the meaning
of the densities in Table 6 if the actual density distribution is
a power law. If
$n(r) = n_1 r_{pc}^{-\alpha}$, with $n_1$ (the density at 1 pc) set
by matching the line profiles (Carr et al. 1995),
the density in Table 6 is reached at
radii of 18\as\ to 7\as\ for $1 \leq \alpha \leq 2$, corresponding to filling
factors of 0.3 to 0.6 in our largest beam. We conclude that, in this source,
the densities derived in this study characterize gas on scales somewhat
smaller than our beams, if the source has a density gradient. Similar studies
of other sources are needed to see if this conclusion can be generalized.

Further evidence for a range of densities is that J=10\to9 emission has
been seen in a number of sources (Hauschildt et al. 1993 and our Table 4).
The data do not warrant detailed source-by-source modeling, but we have
predicted the expected J=10\to9 emission from a source with the mean
properties found in \S 4.1: log(n) = 5.93 and log(N) = 14.42. We assumed
a linewidth of 5.0 \kms, about the mean for our sample, and \tk = 50 K.
The predicted T$_R$ of the J=10\to9 line is 0.2 K for this average cloud,
weaker than any of the detections.
If we use the conditions for the cloud with properties at the high end of the
1 $\sigma$ spread, we can produce T$_R$ = 1.6 K, about the weakest detection.
Increasing \tk\ to 100 K raises the prediction to 7 K, similar to many of
the detections. Detection of a J=10\to9 line therefore implies a cloud with
higher density, column density, and/or temperature than the average cloud
in our sample of sources detected at J=7\to6.

\subsection{Masses }

Table 7 contains mass estimates for the regions for which
we have determined cloud sizes. We have computed three different
estimates.  The first estimate assumes that the volume density
fills a spherical volume with the diameter of the J=5\to4 emission:
\begin{equation}
M_n = {{4}\over{3}}\pi{r^3}{n}{\mu},
\end{equation}
where r is the radius of the cloud
and $\mu=2.34m_H$ is the mean mass per particle.
The second estimate uses the CS column densities (N) and the formula:
\begin{equation}
M_N = \pi{r^2}{{N}\over{X}}{\mu},
\end{equation}
where X is the abundance of CS.  We have used $X = 4 \times 10^{-10}$,
based on a more detailed analysis of one of the sources in this
study (Carr et al. 1995).
Finally, we estimated masses from the virial theorem:
\begin{equation}
M_{V} = {{5}\over{3}}{{R V^2_{rms}}\over{G}},
\end{equation}
for a spherical, non-rotating cloud.  Assuming
that the velocity profile is Gaussian, $V_{rms}$
is related to the FWHM  ($\Delta{v}$) of the line by
$V_{rms} = \sqrt{3} \Delta{v}/2.35$.
We used the average $\Delta{v}$ of the CS lines.
The value of $M_n$ for GL490 is probably underestimated
substantially because the maser position is quite far from
the peak of a very compact source. Zhou et al. (1996) have
analyzed this source in more detail and found considerably higher
densities from spectra on the peak. Consequently, we ignore this
source in the following discussion.

The average ratio of $M_N/M_n$ is $0.84\pm 0.73$.
The agreement is gratifying, but the poorly known abundance of CS makes $M_N$
quite uncertain. In contrast, the agreement between $M_n$ and $M_V$ is worse,
with $M_n$ almost always considerably larger than $M_V$.
A likely explanation is that the gas is distributed inhomogeneously
within the beam, whereas the calculation of
$M_n$ assumes that the density is uniformly distributed.
We have used the ratio of $M_V$ to $M_n$ to estimate the volume
filling factor ($f_v$) of the gas, also listed in Table 7.
The filling factors have a large range (0.02 to 2.3) and a mean
value of $0.33\pm 0.59$.
The virial mass estimate is susceptible to
error because the linewidth may be affected by unbound
motions, such as outflows, and it ignores effects of external
pressure. Least certain is $M_n$, which depends on the cube
of the size (and hence distance). Each mass estimate depends on
a different power of the size, making their ratio strongly dependent
on uncertainties in the distance.
In view of the  problems inherent in each of the
different mass calculations, the masses agree reasonably well.
Because the virial mass estimates have the fewest potential
problems, we will use them in what follows.
The average $M_V = 3800$ M$_{\sun}$.

\section{Implications }

\subsection{Comparison to Other Star-Formation Regions }

Are the high densities seen in this survey
peculiar to regions of massive star formation or are they
a feature of star formation in general?
Lada, Evans, \& Falgarone (1996) have found that
the density in the most active star-forming cores in L1630
is about log(n) = 5.8, very similar to what we find.
We also compared the results of our
study with surveys of regions forming low-mass stars.
\markcite{Zhou et al. (1989)}
observed a sample of low-mass cores in  CS transitions up to J=5\to4
and derived densities of \meann $ =  5.3\pm 1.1$.
These densities are about a factor of 4 lower than the densities
we find in this study (and in other studies of regions of massive
star formation). Since Zhou et al. (1989) did not have J=7\to6 data,
it may be more appropriate to compare with our fits to sources without
J=7\to6 detections; in that case, our densities are larger by
a factor of about 2. The net result is that regions forming massive
stars do seem to have larger densities when similar techniques are
used, but the difference is not an order of magnitude.

The ability to form low-mass stars in regions of massive star formation
may depend on whether the Jeans mass remains low as the cloud is heated.
We can calculate the Jeans mass from
\begin{equation}
 M_J(\rm{M}_{\sun}) = 18T^{{3}\over{2}}n^{-{{1}\over{2}}}.
\end{equation}
Using the mean logarithmic densities and the assumed temperatures
(10 K for the low-mass cores, 50 K for our sample),
we compute  $\langle M_J\rangle = 1.3$ M$_{\sun}$
for the clouds forming low-mass stars
and $\langle M_J\rangle = 7 $M$_{\sun}$ for clouds in this
study with J=7\to6 emission.
The assumed temperatures make $M_J$ higher in regions forming massive stars
even though they are denser. However, the strong dependence of $M_J$
on temperature means that statements about average properties
should not be taken too literally until the temperatures are
known better. In addition, the
fragmentation spectrum may have been established early in the evolution
of the core, before the temperatures were raised by the formation
of massive stars.

\subsection {Do Larson's Laws Apply to Massive Cores?}

Most studies of the global properties of molecular clouds deal with
the usual linewidth--size--density relations, as proposed by Larson (1981)
and confirmed by others (e.g., Fuller \& Myers 1992; Solomon et al. 1987;
Caselli \& Myers 1995). These relations were generally found by comparing
properties of whole clouds; similar relations were found within single
clouds by comparing map sizes in transitions of different molecules.
A recent paper by Caselli \& Myers (1995) includes
information on both low mass cores and more massive cores within
the Orion molecular cloud. They fit the non-thermal linewidth (the
observed linewidth after correction for the thermal contribution) and cloud
radius for these types of regions separately to this relation:
\begin{equation}
{\rm log} \Delta v ({\rm km s^{-1}}) = b + q {\rm log} R({\rm pc}).
\end{equation}
They found a strong relationship (correlation coefficient, $r = 0.81$)
in low-mass cores with $b= 0.18 \pm 0.06$ and $q= 0.53 \pm 0.07$.
The relation was considerably weaker ($r = 0.56$) and
flatter ($q = 0.21 \pm 0.03$) in the massive cores.
In Figure 5, we plot log($\Delta v$) versus log$R$ for the sources
in Table 5, which are generally denser and more massive than the
cores studied by Caselli \& Myers. No relationship is apparent
(the correlation coefficient is only $r = 0.26$),
despite the fact that our sample covers a range of 30 in source size.
Nevertheless, we fitted the data to equation 7 using least squares
and considering uncertainties in both variables (we assumed 20\%
uncertainties in size and used the standard deviation of the linewidths
of the different lines for the uncertainty in $\Delta v$).
The result was $b = 0.92 \pm 0.02$ and $q= 0.35\pm 0.03$, but
the goodness of fit parameter, $Q = 2.8 \times 10^{-8}$,
whereas a decent fit should have $Q>0.001$. Alternatively, we minimized
the mean absolute deviation (robust estimation; see Press et al. 1992).
The result was $b = 0.80$ and $q= 0.08$, indicating essentially no
size--linewidth relation.

Thus our data confirm
the trend discernable in the analysis of Caselli \& Myers: the
$\Delta v - R$ relation tends to break down in more massive cores.
We have plotted the Caselli \& Myers relations in Figure 5, along
with Larson's original relation.
It is clear that our sources have systematically higher linewidths
at a given radius than sources in other studies. For the radii we
are probing, most other studies were considering linewidths from CO or
its isotopes and may thus have included a larger contribution from low-density
envelopes. The usual relations would predict larger $\Delta v$ in these
regions,
which would make the discrepancy worse. However, our sources are regions of
massive star formation,
and Larson (1981) noted that such regions (Orion and M17 in his study) tended
to have larger $\Delta v$ for given size and not to show a size--linewidth
correlation.

Most previous studies have found an inverse relation between $mean$ density
and size, corresponding to a constant column density. However, Scalo (1990)
and Kegel (1989) have
noted that selection effects and limited dynamic range may have produced this
effect, and Leisawitz (1990) found no relationship between density and size
in his study of clouds around open clusters. In previous studies,
the $mean$ densities were found by dividing a column density by a size,
which might be expected to introduce an inverse correlation if the column
density tracer has a limited dynamic range. Since our
densities were derived from an excitation analysis, it may be interesting to
see if any correlation exists in our data.
We plot log(n) versus log$R$ in Figure 5. Again, no correlation
is evident ($r = -0.25$), and our densities all lie well above (factors of
100!)
predictions from previous relations (e.g., Myers 1985). Again, Larson
(1981) noted a similar, though much less dramatic, tendency for regions
of massive star formation in his analysis. For a recent theoretical
discussion of these relations, see V\'azquez-Semadeni, Ballesteros-Paredes,
\& Rodr\'iguez (1997).

To use data on sources without
size information, we plot (in the bottom panel of Figure 5)
log($\Delta v$) versus log(n). The previous relations would
predict a negative slope (typically $-0.5$) in this relation.
In contrast to the predictions, our data show a positive, but small,
correlation coefficient ($r= 0.40$). The slope from a least squares fit
is quite steep ($1.3\pm 0.2$), but robust estimation gives a slope of
only 0.39. In addition, the linewidths are much larger than would have
been predicted for these densities from previous relations.
These results suggests that an uncritical application of scaling
relations based on {\it mean} densities to actual densities, especially in
regions of massive star formation, is likely to lead to errors.

The fact that Larson's laws are not apparent in our data indicate
that conditions in these very dense cores with massive star formation
are very different from those in more local regions of less massive
star formation. The linewidths may
have been affected by star formation (outflows, expanding HII regions,
etc.); the higher densities are probably caused by gravitational
contraction, which will also increase the linewidths.
While the regions in this study may not be typical of most molecular
gas, they are typical of regions forming most of the massive stars
in the Galaxy. These conditions (denser, more turbulent than usually
assumed) may be the ones relevant for considerations of initial mass
functions.

\subsection{Luminosity, Star Formation Efficiency, and Gas Depletion Time}

We have collected data from the literature (or our own
unpublished data) on the luminosity of the sources in Table 7.
The ratio of the luminosity to the virial mass
($L/M$), roughly proportional to the star formation rate per
unit mass, ranges from 24 to 490 in solar units (see Table 7)
with a mean of $190 \pm 43$, where 43 is the standard deviation of the mean
(all other uncertainties quoted in the text are standard
deviations of the distribution).  Previous studies, using masses
determined from CO luminosity, have
found much lower average values of $L/M$: 4.0 for the inner galaxy
(Mooney \& Solomon 1988); 1.7 for the outer galaxy (Mead,
Kutner, \& Evans 1990). In fact, the maximum values in those
samples were 18 and 5, respectively, smaller than any of our
values. The enormous difference is caused by
the fact that we are calculating the mass of the dense gas,
which is much less than the mass computed from the CO luminosity.
While we have also tried to use luminosities measured with
small beams, the main difference is in the mass. One
way to interpret this result is that the star formation rate
per unit mass
rises dramatically (a factor of 50) in the part of the cloud with dense gas.

The star formation rate per unit mass of very dense gas may be more relevant
since stars do not seem to
form randomly throughout molecular clouds (Lada et al. 1991).
Instead, the 4 most massive CS cores in L1630,
which cover only 18\% of the surveyed area,
contain 58\% to 98\% of all the forming stars, depending on
background correction.  Li et al. (1996) have found that there
is little evidence for any recent star formation outside the clusters,
suggesting that the 98\% number is closer to correct.
The star formation efficiency in the clusters can be
quite high (e.g., 40\%) compared to that of the cloud
as a whole (4\%) (Lada et al. 1991).

The gas depletion time ($\tau$) is the time to turn all the molecular gas into
stars.   Considering only stars of $M > 2$ M$_{\sun}$, the star
formation rate can be written as
$dM/dt$ (M$_{\sun}$ yr$^{-1}$) = $4 \times 10^{-10} L$
(Gallagher \& Hunter 1986; Hunter et al. 1986). The coefficient differs
by only 20\% if the lower mass cutoff is 10 M$_{\sun}$.
The gas depletion time can then be written as
$\tau\ = 2.5 \times 10^{9} (M/L)$ yr. Using our value of average $L/M = 190$,
$\tau = 1.3 \times 10^7$ yr. This time is comparable to that for
dispersal of clouds surrounding open clusters; clusters with ages
in excess of $1.0\times 10^7$ yr do not have associated molecular clouds
with masses as large as $10^3 M_{\sun}$ (Leisawitz, Bash, \& Thaddeus
1989).

\subsection{ Luminosity of the Galaxy in CS J = 5\to4}

CS J = 5\to4 emission has been seen toward the centers of
NGC 253, M82, IC 342, Maffei 2, and NGC 6946
(\markcite{Mauersberger and Henkel 1989; }
\markcite{Mauersberger et al. 1989)}.
For comparison to studies of other galaxies, we will estimate the luminosity
of the Milky Way in CS 5\to4 [$L_G ({\rm CS \ 5-4})$] from the mean \lcs\
per cloud in Table 5
and an estimate of the number of such clouds ($n_{cl}$) in the Galaxy.
{}From Table 5
we find $\langle L({\rm CS \ 5-4}) \rangle = 4 \times 10^{-2}$ L$_{\sun}$ and
$\langle\int{T_R^*dv}\rangle = 34$ K km s$^{-1}$, whereas
$\langle\int{T_R^*dv}\rangle = 42$ K km s$^{-1}$ for the whole sample in
Table 2. If we correct for the fact that the mean integrated intensity
of the sources in Table 5 is less than the mean of the whole sample,
we would get $5 \times 10^{-2}$ L$_{\sun}$ for the typical core.

We do not have
a direct measurement of  $n_{cl}$ because our survey is incomplete.
The most recent update to the \water\ maser catalog (Brand et al. 1994)
brings the total
number of masers with IRAS colors characteristic of star formation regions
(see Palagi et al. 1993) to 414.  If we assume
that our CS 5\to4 detection rate of 75\% applies equally to the other
sources, we would expect 311 regions of CS J = 5\to4 emission
in a region which covers two thirds of the galaxy.  If we correct for the
unsurveyed third of the galaxy, we would estimate the total number
of cloud cores emitting CS J =5\to4 to be 466.

Consequently, we will assume $n_{cl} = 311 - 466$,
with the larger values probably being more likely.
Using these numbers, we calculate $L_G ({\rm CS \ 5-4}) = 15 - 23$ L$_{\sun}$.
Even though we have made some completeness corrections, we expect these
estimates to be underestimates because of our limited sensitivity and the
likelihood of CS emission from dense regions without \water\ masers.

These values can be compared with the luminosities of other galaxies
in Table 8. However, our estimate applies to the entire Galaxy excluding
the inner 400 pc,
while the \lcs\ for other galaxies are derived from
a single beam, centered on the nucleus, with a radius given in the
Table. The inner 100 pc of M82 and NGC 253 emit more CS J = 5\to4
than does our entire Galaxy, excluding the inner 400 pc.

We can also compare our Galaxy to others in terms of its star formation rate
per unit mass. In \S 5.3, we used $L/M$, with $M$ as the virial mass, to
measure this quantity. Because linewidths in galaxy observations are likely to
reflect the total mass, rather than the gaseous mass, we will use
$L$/ \lcs\ as a stand-in for the star formation rate per unit mass of dense
gas.
We have tabulated the far-infrared luminosity of the galaxies in Table 8,
using the data with the smallest available beam, to provide the best
match to the CS J = 5\to4 observations, which were mostly done with the
IRAM 30-m telescope ($11\arcsec$ beam).
The resulting $L$/\lcs\ values range from $5.0 \times 10^7$ (NGC 253) to
$1.7 \times 10^9$ (M82). These numbers apply to regions typically 100 pc in
radius.
For our Galaxy, we have only the total \lcs, so we compare to the total
$L = 1.8 \times 10^{10}$ L$_{\sun}$ (Wright et al. 1991). The result is
$8 - 13 \times 10^8$, nominally similar to M82; however, much of the
far-infrared emission of our Galaxy is likely to result from heating by
older stars.
Probably a more useful comparison is to the values of $L$/\lcs\ in individual
clouds (Table 7). No individual cloud approaches the value in M82.
The highest value in Table 7 is about twice that of NGC 253 and half that
of IC 342.

\section{Summary}

\begin{enumerate}

\item Very dense gas is common in regions of massive star formation.
The gas density for the regions selected by having a water maser
is \meann $ = 5.93$ and the CS column density is \meancd $ = 14.42$.
For regions without CS J = 7\to6 emission
the mean density is half as large and the mean column
density is about 7 times smaller.
These results are relatively insensitive to both CS optical
depth and to changes in the kinetic temperature of the region.
The mean density is an order of magnitude less than the critical
density of the J = 7\to6 line because of trapping and multilevel excitation
effects.

\item
In many regions forming massive stars, the
CS emission is well modeled by a single density gas component, but
many sources also show evidence for a range of densities.
{}From simulations of emission from gas composed of two different
densities ($10^4$ and $10^8$ \cmv), we conclude that there are
few clouds with filling factors of ultra-dense gas (n$ = 10^8$ \cmv)
exceeding 0.25.

\item
The densities calculated for the sources in this survey are comparable
to the densities seen from detailed studies of a few individual
regions forming massive stars.
Therefore, it is likely that very dense gas is
a general property of such regions.
The average density of regions forming massive stars
is at least twice the average in regions forming only low-mass
stars.

\item
Using a subsample of  sources whose CS 5\to4 emission was
mapped at the CSO,  the average cloud diameter is 1.0 pc and the average
virial mass is 3800 M$_{\sun}$.

\item
We see no evidence for a correlation between linewidth and size or density
and size in our sample.
Our linewidths and densities are systematically larger at a given size
than those predicted by previous relations.
There is, however, a positive correlation between linewidth and density,
the opposite of predictions based on the usual arguments.

\item
The ratio $L/M$, which is a measure of star formation
rate per unit mass for the dense gas probed by CS J=5\to4 emission,
 ranges from 24 to 490, with an average value of 190.

\item
The dense gas depletion time, $\tau \sim 1.3 \times 10^7$ yr,
comparable to the dispersal time of gas around
clusters and OB associations.

\item
The estimated Galactic luminosity in the CS J = 5\to4 line is
$14-23$ L$_{\sun}$. This range of values is considerably less than
what is seen in the inner 100 pc of starburst galaxies. In addition,
those galaxies have a higher ratio of far-infrared luminosity to
CS J = 5\to4 luminosity than any cloud in our sample.

\end{enumerate}

\acknowledgements

We are grateful to the staff of the IRAM 30-m telescope for assistance
with the observations. We also thank T. Xie, C. M. Walmsley, and J. Scalo
for helpful discussion.
This research was supported in part by NSF Grant AST-9317567 to the
University of Texas at Austin.

\clearpage

\begin{table}[h]
\caption{Observing Parameters}
\vspace {3mm}
\begin{tabular}{l c c c c  c c c }
\tableline
\tableline
Line & $\nu$ & Telescope & $\eta_{mb}^a$ & $\theta_b^a$ & $\langle
T_{sys}\rangle^b$ & $\delta v$ & $\delta v$ \cr
& (GHz)  & &  & ($\arcsec$)  &(K) & (km s$^{-1}$) &(km s$^{-1}$)  \\ \tableline

CS 2\to1        & 97.980968 &  IRAM &   0.60 &   25\as &     675 &    0.31$^c$
&    3.06$^d$ \cr
CS 3\to2        & 146.969049 & IRAM &   0.60 &   17\as &    990 &    0.32$^e$ &
   2.04$^d$ \cr
CS 5\to4        & 244.935606 & IRAM &   0.45 &   10\as &    2500 &    1.22$^f$
&    \nodata \cr
C$^{34}$S 2\to1 & 96.412982 &  IRAM &  0.60 &   25\as &     620  &
0.31$^{c,g}$ &  3.11$^d$ \cr
C$^{34}$S 3\to2 & 144.617147 & IRAM &   0.60 &   17\as &    835 &
0.32$^{e,h}$ &  2.07$^d$ \cr
C$^{34}$S 5\to4 & 241.016176 & IRAM &   0.45 &   10\as &    2700 &    1.24$^f$
&     \nodata \cr
CS 5\to4        & 244.935606 & CSO  &   0.71 &   30\as &    445   &    0.17$^i$
&    1.2$^j$ \cr
C$^{34}$S 7\to6 & 337.396602 & CSO  &   0.55 &   20\as &    1000  &    0.12$^i$
&    0.89$^j$ \cr
CS 10\to9       & 489.75104  & CSO  &   0.39 &   14\as &    4300 &    0.09$^i$
&    0.61$^j$ \cr
CS 14\to13      & 685.434764 & CSO  &   0.31 &   11\as &    2050 &    0.06$^i$
&    0.44$^j$ \cr
\end{tabular}
\tablecomments{(a) Efficiency and beam size; (b) average $T_{sys}$ during
observing;
(c) 100 kHz filterbank; (d) Split 1 MHz filterbank; (e) Autocorrelator; (f) 1
MHz filterbank;
(g) $\Delta{V} = 0.486$ \kms\ for C$^{34}$S 2-1 in autocorrelator;
(h) $\Delta{V} = 0.207$ \kms\ for C$^{34}$S 3-2 in 100 KHz filterbank; (i) 50
MHz AOS;
(j) 500 MHz AOS.}

\end{table}

\clearpage

\begin{table}[h]
\caption{Standin for table 2 ps file.  Discard this page.}
\vspace {3mm}
\begin{tabular}{l r r r r r  }
\tableline
\tableline
Source &  $\int$T$_{R^*}$dV & $V_{LSR}$ & FWHM & $T_R^*(10-9)$ & $T_R^*$(14-13)
\cr
&     K km s$^{-1}$ & km s$^{-1}$ & km s$^{-1}$ & K  & K \\ \tableline

\end{tabular}
\tablerefs{
(a) Carr et al. (1995)}
\end{table}
\clearpage

\begin{table}[h]
\caption{Standin for table 3 ps file.  Discard this page.}
\vspace {3mm}
\begin{tabular}{l r r r r r  }
\tableline
\tableline
Source &  $\int$T$_{R^*}$dV & $V_{LSR}$ & FWHM & $T_R^*(10-9)$ & $T_R^*$(14-13)
\cr
&     K km s$^{-1}$ & km s$^{-1}$ & km s$^{-1}$ & K  & K \\ \tableline

\end{tabular}
\tablerefs{
(a) Carr et al. (1995)}
\end{table}
\clearpage

\begin{table}[h]
\caption{Results for CS $J = 10\rightarrow 9$ and $J= 14\rightarrow 13$ Lines}
\vspace {3mm}
\begin{tabular}{l c c c c c  }
\tableline
\tableline
Source &  $\int$T$_{R^*}$dV & $V_{LSR}$ & FWHM & $T_R^*(10-9)$ & $T_R^*$(14-13)
\cr
& (K km s$^{-1}$) & (km s$^{-1}$) & (km s$^{-1}$) & (K)  & (K) \\ \tableline
GL2591$^a$ &   2.7          & -5.3      & 1.6    & 1.6     &         \cr
S158A      &   22.6         & -57.2     & 2.9    & 7.2     & \nodata \cr
W3(2)      &  6.4           & -38.49    & 2.28   &  2.6    & \nodata \cr
W3(OH)     &  \nodata       & \nodata   &\nodata & \nodata &  $<1.6$ \cr
S255       &   10.3         & 8.2       & 2.3    & 4.4     &  $<0.7$ \cr

\end{tabular}
\tablerefs{
(a) Carr et al. (1995)}
\end{table}
\clearpage

\begin{table}[h]
\caption{Diameters, Offsets and Luminosities from CS J = 5$\rightarrow$4 Maps}
\vspace {3mm}
\begin{tabular}{l c c c c c c c }
\tableline
\tableline
Source & ref & Dist. & $\int$T$_{R^*}$dV & \lcs\  & Beam Corr. & Diameter &
Offset \cr
&   & (kpc) &  (K km s$^{-1}$) & (10$^{-2}$ L$_{\sun}$) & & (pc) & (arcsec) \\
\tableline
W43S & & 8.5 & 52.8 & 6.1 & 1.5 & 0.9 & (0,5) \cr
W43Main1 & & 7.5 & 22.1 & 5.2 & 4.0& 1.9 & (20,-36) \cr
W43Main3 & & 6.8 & 32.4 & 4.6 & 2.9 & 1.4 & (-8,2) \cr
31.25-0.11 & & 13 & 9.0 & 5.7 & 3.6 & 3.0 & (-12,-15) \cr
31.44-0.26 & & 9.4 & 23.0 & 8.6 & 4.0 & 2.4 & (-2,-4) \cr
32.8+0.2A & & 15 & 64.1 & 15 & 1.0 & $<$1.1 & (-5,-4) \cr
W44 & & 3.7 & 87.9 & 3.1 & 2.5 & 0.7 & (-3,0) \cr
W51W & & 7 & 12.0 & 1.6 & 2.6 & 1.3 & (0,-7) \cr
W51N & & 7 & 79.3 & 17 & 4.2 & 1.8 & (0,-5) \cr
W51M & & 7 & 152 & 19 & 2.4 & 1.2 & (-3,-2) \cr
ON1 & & 6 & 24.4 & 1.6 & 1.7 & 0.7 & (0,0) \cr
K3-50 & & 9 & 11.3 & 1.9 & 2.0 & 1.3 & (-5,5) \cr
ON3 & & 9 & 11.0 & 1.8 & 2.0 & 1.3 & (0,-4) \cr
ON2S & & 5.5 & 22.3 & 1.5 & 2.2 & 0.9 & (-6,0) \cr
ON2N & & 5.5 & 15.4 & 1.0 & 2.1 & 0.8 & (6,5) \cr
S106 & & 0.6 & 5.4 & 0.004 & 2.2 & 0.1 & (20,0) \cr
CRL 2591 & 1& 1.0 & 7.9 & 0.024 & 3.3 & 0.22 & (0,0) \cr
DR21 S & & 3 & 44.8 & 1.0 & 2.3 & 0.5 & (-6,5) \cr
W75(OH) & & 3 & 47.6 & 1.1 & 2.4 & 0.5 & (-6,-5) \cr
W75S1 & & 3 & 9.4 & 0.9 & 9.7 & 1.3 & (-3,7) \cr
W75S3 & & 3 & 6.8 & 0.2 & 3.2 & 0.7 & (0,0) \cr
W75N & & 3 & 35.2 & 0.8 & 2.5 & 0.5 & (-5,6) \cr
CepA & 2 & 0.73 & 30.0 & 0.1 & 5.5 & 0.2 & (10,12) \cr
W3(2) & 2 & 2.3 & 26.3 & 0.8 & 5.5 & 0.7 & (0,12) \cr
GL 490 & 2 & 0.9 & 7.5 & 0.01 & 1.8 & 0.12 & (-14,-12) \cr
\end{tabular}
\tablerefs{
(1) Carr et al. (1995); (2) Zhou et al. (1996)}
\end{table}
\clearpage

\begin{table}[h]
\caption{Standin for table 6 ps file.  Discard this page.}
\vspace {3mm}
\begin{tabular}{l r r r r r  }
\tableline
\tableline
Source &  $\int$T$_{R^*}$dV & $V_{LSR}$ & FWHM & $T_R^*(10-9)$ & $T_R^*$(14-13)
\cr
&     K km s$^{-1}$ & km s$^{-1}$ & km s$^{-1}$ & K  & K \\ \tableline

\end{tabular}
\tablerefs{
(a) Carr et al. (1995)}
\end{table}
\clearpage

\begin{table}[h]
\caption{Masses and Luminosities}
\vspace {3mm}
\begin{tabular}{l c c c c c c c c c}
\tableline
\tableline
Source & Flag & $M_n$ & $M_N$ & $M_V$ & $f_v$  & $L$ & Ref. & $L/M_V$ & $L$/
\lcs\ \cr
 & & ($M_{\sun}$) & ($M_{\sun}$) & ($M_{\sun}$) &  & ($L_{\sun}$) &
&($L_{\sun}/M_{\sun}$) &($10^7$) \\ \tableline
W43 S& C$^{34}$S& $2.3\times10^4$& $2.8\times10^4$& $1.8\times10^3$& 0.08&
\nodata & \nodata & \nodata & \nodata\cr
31.44$-$0.26& C$^{34}$S& $3.9\times10^5$& $1.2\times10^5$& $6.3\times10^3$&
0.02& \nodata & \nodata & \nodata & \nodata \cr
32.8+0.20 A& C$^{34}$S& $5.6\times10^4$& $1.3\times10^4$& $7.0\times10^3$&
0.13&  \nodata & \nodata & \nodata & \nodata \cr
W44 & C$^{34}$S& $1.6\times10^4$& $3.9\times10^4$& $4.5\times10^3$& 0.27&
$3.0\times10^5$ & 4 & 67 & 1.0 \cr
W51 W& C$^{34}$S& $5.9\times10^4$& $1.4\times10^4$& $1.5\times10^3$& 0.03&
\nodata & \nodata & \nodata & \nodata \cr
W51 N C2& C$^{34}$S& $6.2\times10^5$& $1.2\times10^5$& $1.3\times10^4$& 0.02&
$4.0\times10^6$ & 4 & 310 & 2.4 \cr
W51M& CS & $7.2\times10^4$& $8.8\times10^4$& $1.6\times10^4$& 0.23&
$2.8\times10^6$& 3 & 170 & 1.5 \cr
K3$-$50& C$^{34}$S& $5.9\times10^4$& $9.4\times10^3$& $6.1\times10^3$& 0.10&
$2.1\times10^6$& 5 & 340 & 11 \cr
ON3& C$^{34}$S& $1.7\times10^4$& $7.3\times10^3$& $2.3\times10^3$& 0.13&
$3.7\times10^5$& 5 & 160 & 2.1 \cr
ON2S&C$^{34}$S& $3.3\times10^4$& $8.0\times10^3$& $9.1\times10^2$& 0.03 &
\nodata & \nodata & \nodata & \nodata \cr
CRL2591& CS& $3.0\times10^2$& $5.0\times10^2$& $3.2\times10^2$& 1.1&
$2.0\times10^4$& 2&  63 & 8.3 \cr
DR21 S& C$^{34}$S& $3.6\times10^3$& $3.5\times10^3$& $1.1\times10^3$& 0.31&
$5.0\times10^5$& 6 & 460 & 5.0 \cr
W75(OH)& C$^{34}$S& $5.6\times10^3$& $9.6\times10^3$& $1.6\times10^3$& 0.27&
$5.4\times10^4$& 8 & 35 & 0.5 \cr
W75 N& C$^{34}$S& $6.6\times10^3$& $3.8\times10^3$& $1.4\times10^3$& 0.22&
$1.8\times10^5$& 3 & 130 & 2.3 \cr
Cep A& CS& $2.5\times10^2$& $4.3\times10^2$& $5.9\times10^2$& 2.3&
$2.5\times10^4$& 1 &42 & 2.5 \cr
W3(2)& C$^{34}$S& $1.9\times10^4$& $2.6\times 10^3$& $6.1\times10^2$& 0.03&
$3.0\times10^5$& 7 & 490 & 3.8 \cr
GL490& CS & $6.2$& $2.8\times10^1$& $9.1\times10^1$& 15& $2.2\times10^3$& 2& 24
& 2.2 \cr

\end{tabular}
\tablerefs{
(1) Evans et al. (1981); (2) Mozurkewich et al. (1986); (3) Jaffe, unpublished;
(4) Jaffe et al. (1985);
(5) Thronson \& Harper (1979); (6) Colom\'e et al. (1995); (7) Campbell et al.
(1995)}
\end{table}
\clearpage

\begin{table}[h]
\caption{Comparison to Other Galaxies}
\vspace {3mm}
\begin{tabular}{l c c c c c c c c  }
\tableline
\tableline
Source &  Distance  & Radius &  $\int{T^*_Rdv}$ & \lcs\ & Ref. & $L$ & Ref. &
$L$/ \lcs\ \cr
&   (Mpc) & (pc) & (K km s$^{-1}$) & (L$_{\sun}$)  &  & ($10^9$ L$_{\sun}$) & &
($10^7$) \\ \tableline

NGC 253          & 2.5$^a$ & 67 & 23.5 & 154 & 1 & 8 & 3 & 5 \cr
Maffei 2         & 5 & 133 & $<$2 & $<$53 & 2 & 9.5 & 2 & $>18$ \cr
IC 342           & 1.8$^b$ & 48 & 0.76 & 3 & 1 & 0.64 & 4 & 21 \cr
M82              & 3.2$^c$ & 85 & 2.6 & 28 & 1 & 47 & 3 & 170 \cr
NGC 6946         & 5 & 133 & $<$2.8 & $<$74 & 2 & 1.2 & 3 & $>1.7$ \cr
\end{tabular}

\tablerefs{ (a)  de Vaucouleurs (1978); (b)  McCall (1987); (c) Tammann \&
Sandage (1968);
(1)  Mauersberger \& Henkel (1989); (2)  Mauersberger et al. (1989);
(3) Smith \& Harvey 1996;
(4) Becklin et al. (1980) for flux, McCall (1987) for distance. }
\end{table}

\clearpage
\begin{figure}
\caption { (a)
Example of a source which we assumed was composed of separate
cloud components.  Notice that the smaller spectral feature is
visible in all transitions except for C$^{34}$S 5\to4 and 7\to6.
(b)  Example of a source for which we assumed that
the multiple spectral features were a result of self-absorption.
}
\end{figure}

\begin{figure}
\caption
{ Bar graph illustrating the  fraction of sources which  have T$_R^* \geq
0.5$K.  Each  bar represents a different  CS transition.
The numbers displayed in each of the bars
are the actual number that were detected.
}
\end{figure}

\begin{figure}
\caption
{ The top panel is the distribution of cloud densities as determined from LVG
model fits to multiple CS transitions.
The solid line shows densities determined
from fits to all 4 CS transitions.
The dashed line shows densities determined
from fits to CS 2\to1, 3\to2, and 5\to4
only in sources with undetected CS 7\to6 emission.
Bins are 0.1 wide in the log.
The bottom panel is the distribution of cloud column densities as
determined from LVG model fits to multiple CS transitions.
The solid line shows the column densities determined
from fits to all 4 CS transitions.
The dashed line shows column
densities determined from fits to CS 2\to1, 3\to2, and 5\to4
only in sources with undetected CS 7\to6 emission.
Bins are 0.1 wide in the log.
}
\end{figure}

\begin{figure}
\caption
{ Examples of LVG model fits to the CS line temperatures
for 6 sources.  The filled circles represent the observations and the solid
line plots the line temperatures determined from the LVG model that fitted
the observations best.
The large error bars on the CS 3\to2 data points are a result
of the large calibration uncertainties for this transition.
The dashed lines are the results of fits with two densities, as described in
the text.
}
\end{figure}

\begin{figure}
\caption
{ The top plot is the log of the linewidth versus
the log of the radius, with the best-fitting straight lines shown as a
dash-dotted line (least-squares) and solid line (robust estimation).
The other lines are taken from the relations of
Larson (1981) and Caselli \& Myers (1995).
The middle plot is the log of the density versus the log of the radius.
The dashed line is from Myers (1985), assuming that his ``size" was
a diameter.
The bottom plot is the log of the
CS linewidths (averaged over all available CS lines)
versus the log of the density.  Open circles represent the sources
with detectable CS 7\to6 emission and the open pentagons represent
those that were not detected in 7\to6.
The filled squares are sources with self absorption.
The best-fitting straight line is shown as a solid line (least squares) and
as a dotted line (robust estimation).
}

\end{figure}

\end{document}